\documentclass[journal]{IEEEtran}
\usepackage{graphicx}

\begin{document}
\title{The s-CVD Radiation Monitoring and Beam Abort System of the Belle-II Vertex Detector}

\author{L. Bosisio, C. La Licata, L. Lanceri, L. Vitale
\thanks{Manuscript received December 10, 2016.}
\thanks{L. Bosisio is with Dipartimento di Fisica, Universit\'a di Trieste, I-34127 Trieste, Italy and INFN Sezione di Trieste, I-34127 Trieste, Italy}%
\thanks{C. La Licata is with Dipartimento di Fisica, Universit\'a di Trieste, I-34127 Trieste, Italy and INFN Sezione di Trieste, I-34127 Trieste, Italy}%
\thanks{L. Lanceri is with Dipartimento di Fisica, Universit\'a di Trieste, I-34127 Trieste, Italy and INFN Sezione di Trieste, I-34127 Trieste, Italy}%
\thanks{L. Vitale is with Dipartimento di Fisica, Universit\'a di Trieste, I-34127 Trieste, Italy and INFN Sezione di Trieste, I-34127 Trieste, Italy}%
}

\maketitle
\pagestyle{empty}
\thispagestyle{empty}

\begin{abstract}
The Belle-II VerteX Detector (VXD) is a 6 layers silicon tracker device that will cope with an unprecedented luminosity of 8 $\times$ $10^{35}$ cm$^{-2}$ s$^{-1}$ achievable by the new SuperKEKB e$^{+}$e$^{-}$ collider, now under commissioning at the KEK laboratory (Tsukuba, Japan).
A radiation monitoring and beam abort system has been developed based on single-crystal s-CVD diamond sensors. The sensors will be placed in 20 key positions in the vicinity of the interaction region. The severe space limitations require a remote readout of the sensors.
In this contribution we present the system design, along with the sensor characterisation procedure. We present also the preliminary results with the prototype system during the first SuperKEKB commissioning phase in February-June 2016.
\end{abstract}

\section{Introduction}
\IEEEPARstart {T}{he} Belle II \cite{belle2} detector is currently under construction at the SuperKEKB electron-positron high-luminosity collider. It has been designed in order to probe fundamental physics issues including studies of  the CP violation in the B meson decays and searches for deviations from the Standard Model of particle physics by providing extremely precise measurements of rare particle decays, continuing the investigation done by the Belle experiment \cite{belle} that successfully operated from 1999 to 2010. 
The SuperKEKB collider is the upgrade of the KEKB collider \cite{kekb} at KEK laboratory in Tsukuba (Japan). 
Its design instantaneous luminosity, 40 times higher than that of KEKB, will be achieved by moderately higher beam currents and considerably smaller beam sizes at the interaction point. 
All the Belle II sub-detectors have been redesigned in order to improve their performances with respect to Belle and to cope with unprecedented luminosity. Identification of primary and secondary vertices of particle tracks is needed for full kinematic reconstruction of the event. To reconstruct a collision event, as close to its origin in the interaction point as possible, the Vertex Detector (VXD) is employed. Close to the beam pipe, it has to operate in a very harsh environment that might strongly affect its properties in a time shorter than the expected lifetime of the experiment. As a consequence of the high luminosity, high beam-induced backgrounds and radiation doses are expected. In order to detect beam conditions that could damage the VXD detector and its front-end electronics, a monitoring system \cite{monitoring} \cite{Vitale:Vertex2014} \cite{Vitale:Pisa2015} needs to be developed.

\section{The Radiation Monitoring System}
The Belle II radiation monitoring system will be based on 20 single crystal diamond sensors (with dimensions $4.5 \times 4.5 \times 0.5$ mm$^3$) obtained using the Chemical Vapour Deposition (CVD) technique. The choice of detectors employed in the radiation monitoring system has been guided by the requirement to have radiation hard detectors in order to cope with the high radiation level expected at Belle II experiment.
Compared to any other known solid-state sensor, diamond detectors are able to operate longer in a very harsh environment at high temperatures and under intense irradiation. The radiation and temperature resistance for this type of detector is related to a wide band gap (5.5 eV) high displacement energy (42 eV), and extreme thermal conductivity (2000 W m$^{-1}$ K$^{-1}$). Diamond sensors, tested as beam loss monitors by Belle and BaBar \cite{babar}, have been extensively used by CDF at FermiLab \cite{cdf} and more recently by CMS \cite{cms} and ATLAS \cite{atlas} at LHC in a high radiation environment.

The main goal of the radiation monitoring system is the measurement of both instantaneous and integrated doses in the inner part of the Vertex Detector of Belle II in order to protect it from large beam losses, aborting the electron and positron beams.
Two different conditions have to be faced and so two different reactions are foreseen. One could be a sudden and large increase in backgrounds while the other is a lesser increase that however could bring to an unacceptable integrated dose over some longer period.   
An immediate trigger signal to the SuperKEKB beam-abort system is needed in the first case (”fast” abort trigger) while a warning signal followed after some time by a beam-abort trigger signal is the corresponding action in the second case (”slow” abort trigger). 
The ”fast” abort trigger system, protecting against radiation bursts, should be able to measure instantaneous dose-rates up to about 50 krad/s with a precision of 50 mrad/s, on the time scale set by the beam revolution period of about 10 μs.
The ”slow” abort trigger, protecting against long-term radiation damage, should be able to measure instantaneous rates with an accuracy of about 5 mrad/s, allowing a 10\% accuracy at a dose rate threshold of 50 mrad/s.

\section{Diamond Sensors Characterization}
The diamond sensors response can be influenced by several issues such as presence of defects in the bulk that can act as trapping and recombination centres. Moreover their behaviour could be affected by semiconductor-metal interface properties at the two electrodes that can have an ohmic or blocking nature.
For these reasons, before the installation a complete characterisation of all devices is needed.

\begin{figure}[!htbp]
\centering
\includegraphics[width=3in]{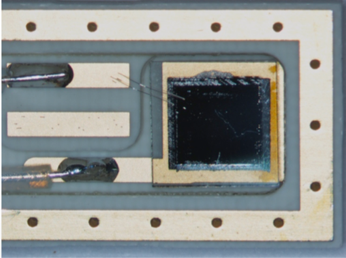}

\caption{A picture of a diamond sensor mounted on a small printed circuit board.}
\label{fig_sensor}
\end{figure}

Four prototypes, three single-crystalline (scCVD) and one poly-crystalline (pcCVD) diamond sensors, have been calibrated at INFN-Trieste laboratories and tested during the first SuperKEKB commissioning phase from February 2016 to June 2016.
All diamonds tested are "Electronic Grade" provided by ElementSix \cite{elementSix}.
Two scCVD and the pcCVD from Micron company are metallised with Al (shadow-mask technique), while the other scCVD from Cividec is metallised with Au/Pt/Ti deposition with thickness 250/120/100 nm respectively.
In fig.\ref{fig_sensor} a scCVD mounted in its compact package is shown.

Before the installation at SuperKEKB for the first commissioning phase, a complete characterisation of the four devices has been performed through several steps:
\begin{itemize}
\item A preliminary test is done measuring the I-V characteristic in the dark and with light exposure. Typical dark currents are $10\div100$ fA for an applied voltage below 200V. The current rises up for applied voltages up to 400V but it still remain under 10pA.  
\item I-V characteristics with the detector irradiated by a $^{90}$Sr $\beta$ source at a fixed distance.
\item Measurement of the current as a function of the relative distance between the source and the sensor and comparison with a FLUKA simulation results. 
\item Long-term stability when irradiated by the $^{90}$Sr $\beta$ source at a fixed voltage and distance between the source and the detector for periods from several hours to few days. 
\item Charge Collection Efficiency (CCE) from measurements with minimum ionising particles.
\item Crystal quality checks and carrier mobilitie measurements exploiting the Transient Current Technique (TCT) with a $\alpha$ source. 
\end{itemize}
 
\section{Measurements with a $\beta$ Source}
\begin{figure}[!htbp]
\centering
\includegraphics[width=3.5in]{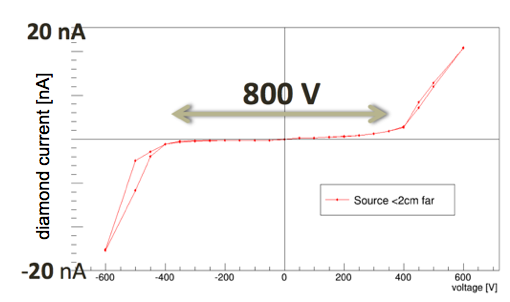}

\caption{The I-V characteristic for a single-crystalline diamond sensor for a reference distance of 1.8cm between the source and the device. A symmetric behaviour is shown for positive and negative voltages applied.}
\label{fig_IVcurve_single}
\end{figure}

\begin{figure}[!htbp]
\centering
\includegraphics[width=3.5in]{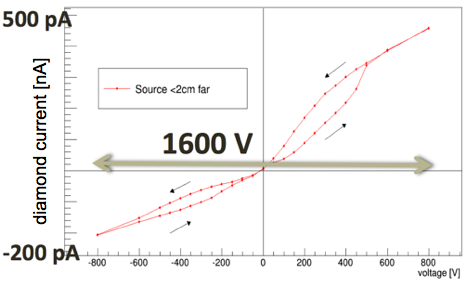}

\caption{The I-V characteristics for a poly-crystalline diamond sensor for a reference distance of 1.8cm between the source and the device. It shows a more pronounced hysteresis behaviour rather than single-crystalline ones}
\label{fig_IVcurve_poly}
\end{figure}

The current-voltage (I-V) characteristic have been performed with the detector irradiated by a $^{90}$Sr $\beta$ source at a fixed reference distance of 1.8 cm. The measurements were performed biasing the detector’s back and reading the signal on the other electrode applying both the polarities. The single-crystalline and the poly-crystalline diamond sensor show in this case a very different behaviour (fig. \ref{fig_IVcurve_single}-\ref{fig_IVcurve_poly}). 
The response of the poly-crystalline diamond device is not reproducible when the applied voltage is increased or decreased. A hysteresis loop forms in the I-V curve, both for positive and negative bias applied, and its presence can be attributed to deep levels traps in the bulk that can act as capture or emission centres of majority carriers during the charging or discharging process. 

\begin{figure}[!htbp]
\centering
\includegraphics[width=3.5in]{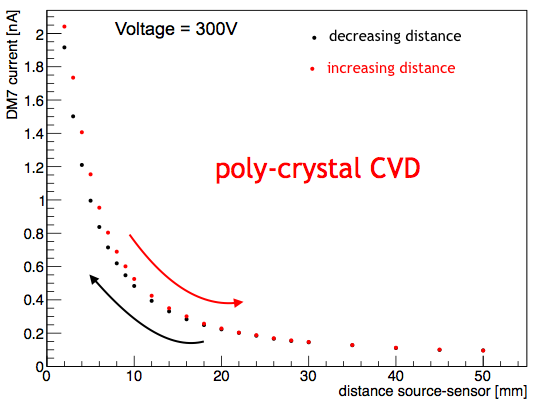}

\caption{Current response of the poly-crystalline diamond sensor as function of the distance between the source and the sensor. The red dots refer to the measurements obtained increasing the distance in the range [2mm, 50mm] while the black dots refer to the current values measured decreasing the distance in the range [50mm, 2mm].}
\label{fig_I_distance}
\end{figure}

\begin{figure}[!htbp]
\centering
\includegraphics[width=3.55in]{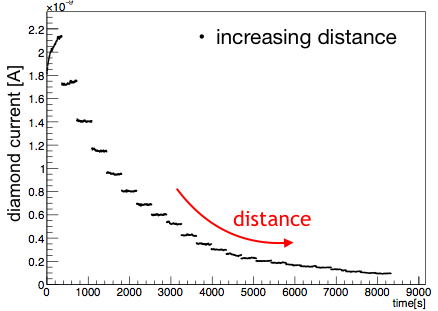}

\caption{Current response of the poly-crystalline diamond sensor as function of time changing the distance between the source and the sensor. In this case the distance is increased from 2mm to 50mm. The first part of the plot refers to measurement taken at 2mm when the sensor does not show a steady response.}
\label{fig_I_distance_increasing}
\end{figure}

\begin{figure}[!htbp]
\centering
\includegraphics[width=3.55in]{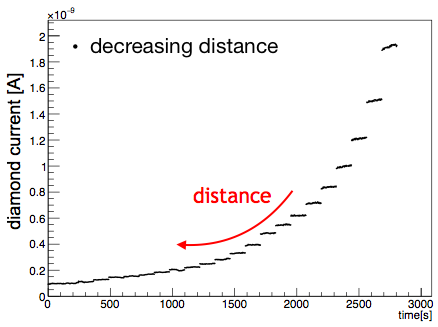}

\caption{Current response of the poly-crystalline diamond sensor as function of time changing the distance between the source and the sensor. In this case the distance is decreased from 50mm to 2mm. The last part of the plot refers to measurement taken at 2mm. In this case the sensor shows a more steady response.}
\label{fig_I_distance_decreasing}
\end{figure}

Another hint of the presence of traps inside the bulk is provided by the measurement of the current as function of the relative distance between the $\beta$ source and the sensor. Different values of the current (fig. \ref{fig_I_distance}) are measured if the distance is increased or decreased especially for very low distances (less than 2 cm).
The coexistence of two different current values at low distances indicates that the values of current are not measured when the sensor is in a steady state. 
Looking at these measurements also as function of time a not stable behaviour is shown when the distance is increased and the source is located at 2mm of distance (fig. \ref{fig_I_distance_increasing}). On the contrary,  a more stable trend is found at 2mm when the distance is decreased (fig. \ref{fig_I_distance_decreasing}). 
 
\begin{figure}[!htbp]
\centering
\includegraphics[width=3.5in]{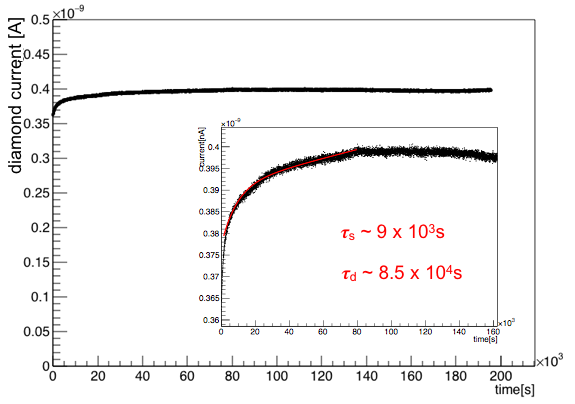}

\caption{The current as function of time for the poly-crystalline diamond sensor. The initial exponential trend is compatible with the Shockley-Hall-Read theory for impurity levels as trapping centres. The picture shows also the fit performed using the function in eq\ref{eq_I_t} and the estimate from the fit of the time constant for shallow and deep traps.}
\label{fig_I_t_withFit}
\end{figure}

Long-term stability when the sensor is irradiated by $\beta$ source at a fixed voltage and distance for periods from several hours to few days is also performed. The fig. \ref{fig_I_t_withFit} shows the current as function of time (I-t curve) obtained for the poly-crystalline diamond sensor. 
It reaches a steady state in about few hours while the observed initial exponential trend of the I-t curve is compatible with the Shockley-Hall-Read theory for impurity levels as trapping centres.
Deep and shallow levels act as recombination centres, thus they must be filled in order to achieve the steady state.
Considering the effect of both deep and shallow traps one can parametrise the current response as:
\begin{equation}
I(t) = I_{0} [1- w_{s}e^{-t/\tau_{s}}-w_{d} e^{-t/\tau_{d}}]
\label{eq_I_t}
\end{equation}
where:
\begin{itemize}
\item $I_{0}$ is the asymptotic value of the current 
\item $w_{s}$ is the weight related to shallow traps
\item $w_{d}$ is the weight related to deep traps
\item $\tau_{s}$ is the time constant associated to shallow traps
\item $\tau_{d}$ is the time constant associated to deep traps
\end{itemize}

\subsection{FLUKA Simulation Comparison}
\begin{figure}[!htbp]
\centering
\includegraphics[width=3.5in]{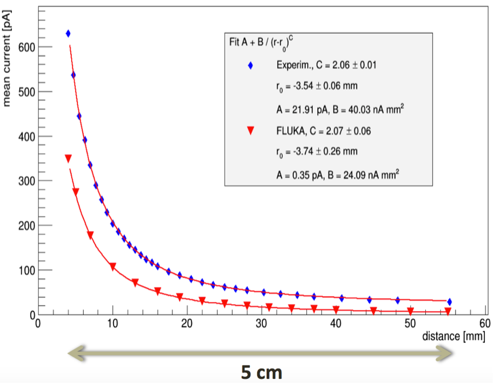}

\caption{The plot show the comparison between the current predicted by FLUKA simulation (red dots) as function of the distance between the source and the sensor and the results from a measurement (blue dots). Both the graphs are fitted using the function: $I(r) = A + B/(r-r_{0})^{C}$ and in both cases the current decrease approximately with the inverse square of the distance.}
\label{fluka}
\end{figure}

The measurements of the current as function of the relative distance between the source and sensors show a dependence on the bias voltage applied.
An explanation on that could be related to the ohmic properties of the diamond-electrode interface.
The contact could be ohmic if a negligible resistance is present at the interface between diamond and electrode and so the charge carriers can be injected from the electrode to the bulk or it can be blocking if a high resistance prevents the flux of charge carriers from one material to the other. 
In order to estimate the photoconductive gain, a comparison with a FLUKA\cite{fluka} simulation results is performed. 
A $4.5 \times 4.5 \times 0.5$ mm$^{3}$ diamond detector fixed on a steel wall and screened by an Al shielding $13 \times 20 \times 0.2$ mm$^{3}$ has been simulated.  
The geometry and the experimental set-up have been implemented using FLUKA simulation and the average energy rate ($\frac{\Delta E}{\Delta t}$), released by electron from a $\beta$ source as a function of the relative distance between the source and the sensor, has been simulated.
The current predicted by FLUKA can be evaluated by the following expression: 
\begin{equation}
I_{FLUKA} = \frac{\Delta E}{\Delta t} \cdot q_{e}\cdot \frac{1}{E_{e-h}} \cdot CCE_{FLUKA} \cdot (G_{PC,FLUKA})
\end{equation}
where $q_{e}$ is the electron charge, $E_{e-h}$ is the average energy needed for an electron-hole pair creation (equal to 13 eV), $CCE_{FLUKA}$ is the charge collection efficiency and $G_{PC,FLUKA}$ is the photoconductive gain. The assumption is that $CCE_{FLUKA}$ and $G_{PC,FLUKA}$ are both equal to 1.\\ 
In fig. \ref{fluka} the measured current and the predicted current by FLUKA as function of the distance are shown. The measured current is greater than the one predicted by the simulation for all the considered distances.

\begin{figure}[!htbp]
\centering
\includegraphics[width=3in]{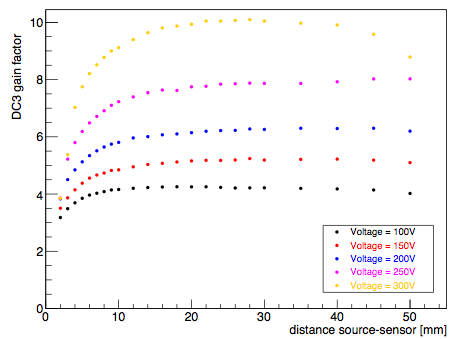}

\caption{The ratio of the measured current and the current from FLUKA simulation for a single-crystalline diamond sensor as function of the distance between the source and the sensor for different applied voltages.}
\label{photoconductive_gain_single}
\end{figure}

\begin{figure}[!htbp]
\centering
\includegraphics[width=3in]{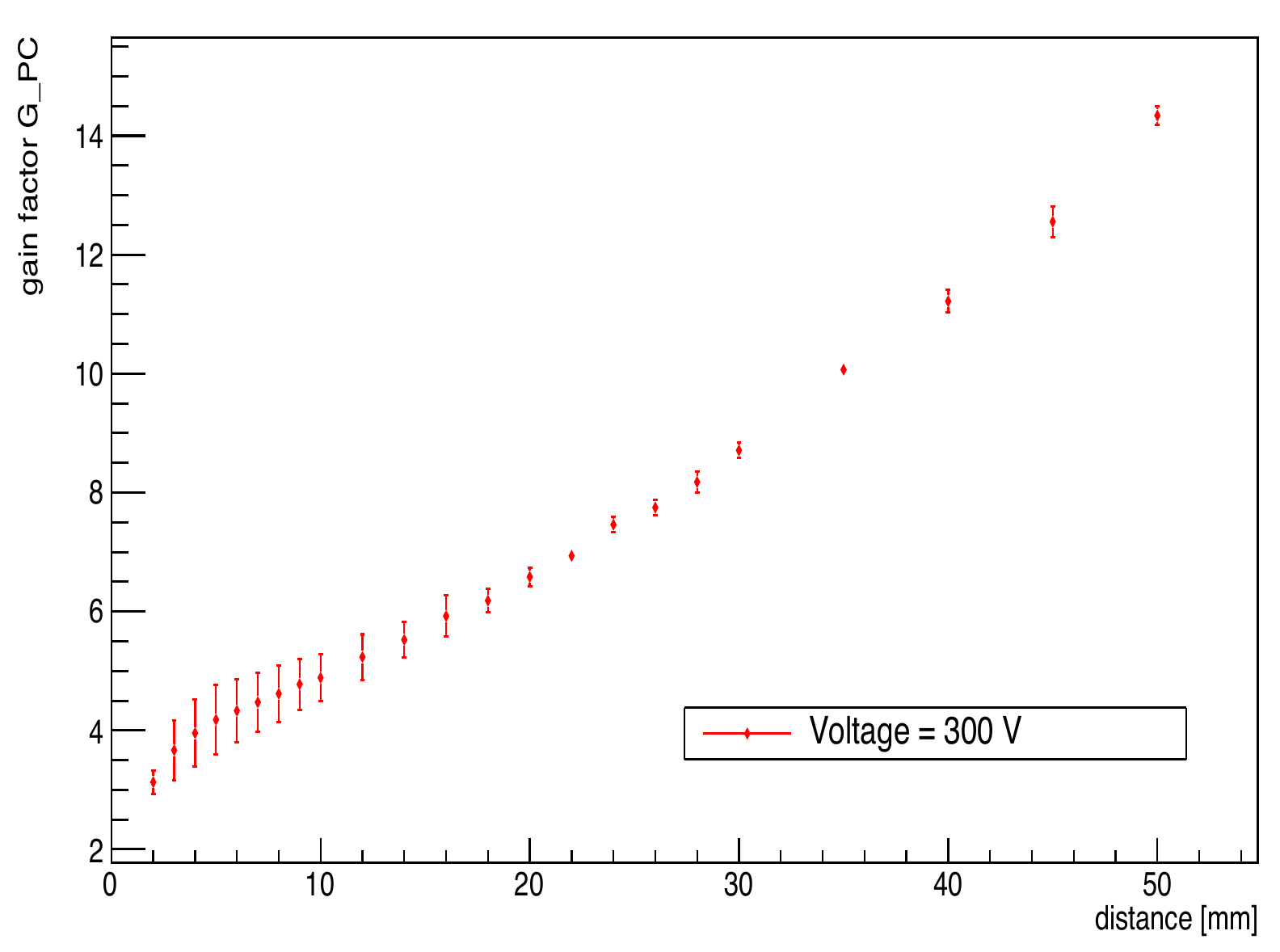}

\caption{The ratio of the measured current and the current from FLUKA simulation for a poly-crystalline diamond sensor as function of the distance between the source and the sensor for an applied voltage of 300V.}
\label{photoconductive_gain_poly}
\end{figure}

Assuming an equal generated charge both for simulation and measurements, by comparing the measured current ($I_{meas}$) and the simulated current ($I_{FLUKA}$) we have an evaluation an overall gain factor:

\begin{equation}
\frac{I_{meas}}{I_{FLUKA}} = \frac{CCE_{meas}\cdot G_{ph,meas}}{CCE_{FLUKA} \cdot G_{ph,meas}} = \frac{CCE_{meas}}{G_{ph,meas}}
\end{equation}

where the $CCE_{meas}$ is the charge collection efficiency of the diamond sensor and $G_{ph,meas}$ its photoconductive gain.
In fig.\ref{photoconductive_gain_single} and fig.\ref{photoconductive_gain_poly} the ratio of the measured current and the current from FLUKA simulation is shown as function of the distance for a single-crystalline and a poly-crystalline diamond sensor respectively. For the single-crystalline diamond sensor a stable gain value is observed for distances greater than 10 mm. The region under 10 mm needs further investigations and the simulation needs to be extended for very small source distances (below 5mm).

\section{Charge Collection Efficiency Measurement}
\begin{figure}[!htbp]
\centering
\includegraphics[width=3.5in]{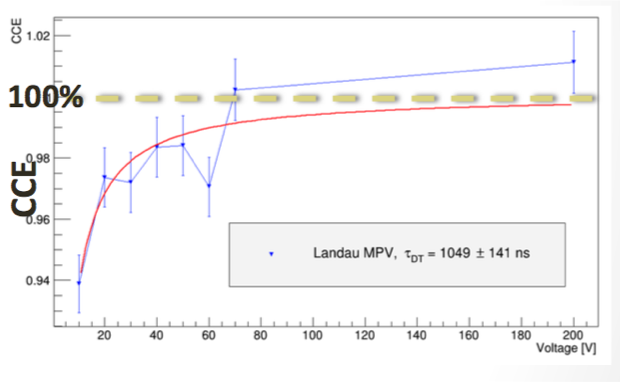}
% where an .eps filename suffix will be assumed under latex, 
% and a .pdf suffix will be assumed for pdflatex; or what has been declared
% via \DeclareGraphicsExtensions.
\caption{The charge collection efficiency measured (blue dots) for a diamond sensor. A $100\%$ charge collection efficiency is reached for applied voltages greater than 70 V. }
\label{CCE}
\end{figure}

For each diamond sensor a measurement of the Charge Collection Efficiency (CCE) has been performed using a beam of minimum ionising electrons from a 90$^Sr$ $\beta$ source. A magnet  selects the electrons of energy around $1.5\div2$ MeV and bends them towards a collimator.
Data proportional to the charge reaching the electrodes is stored. The peak of the Landau distribution gives an estimate of the collected charge.  
In fig.\ref{CCE} the CCE measured as function of the applied voltage is shown. A $100\%$ charge collection efficiency is reached for applied voltages greater than 70 V. 
A fit is performed considering the Hecht's equation:
\begin{equation}
CCE(V) = \frac{Q_{collected}}{Q_{generated}} = \frac{v_{dr}\tau_{DT}}{d_{D}}\cdot (1-e^{-\frac{d_{D}}{v_{dr}\tau_{DT}}})  
\end{equation}
where $v_{dr}$ is the charge drift velocity, $\tau_{DT}$ is the lifetime of carriers due to deep traps recombination, $d_{D}$ is the detector’s thickness.

\section{Transient Current Technique}
\begin{figure}[!t]
\centering
\includegraphics[width=3.5in]{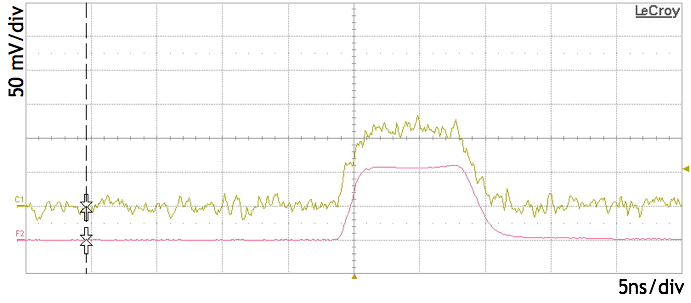}

\caption{Induced current pulse from a TCT measurement. The yellow line is the pulse of a single event while the red line is the average signal obtained from 1000 events.}
\label{TCT_pulse}
\end{figure}

The crystal quality has been checked and charge-transport parameter such as the carriers mobilities measured performing the Transient Current Technique (TCT). 
This technique exploits $\alpha$ particles from a $^{241}$Am radioactive source that releases all the energy within a few microns just below the irradiated electrode. According to the bias polarity, one type of the generated charge carriers is quickly absorbed into the electrode, while the other carrier type moves through the diamond bulk in the direction of the opposite electrode. This method allows for studying separately the drift of electrons and holes in diamond sensors.
The width of the $\alpha$-induced pulse is related to the charge carrier drift time while its shape gives information about the quality of the crystal. Deviations from a rectangular pulse shape can be due to presence of impurities and imperfections inside the bulk.  
As an example fig. \ref{TCT_pulse} shows the induced signal for a single crystal diamond sensor. 

\section{BEAST phase I}
\begin{figure}[!htbp]
\centering
\includegraphics[width=3.5in]{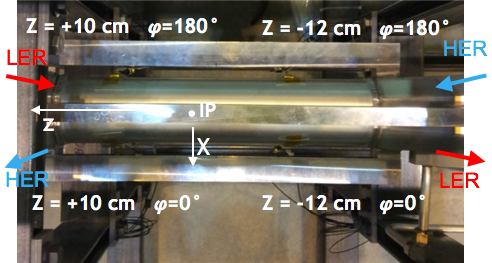}

\caption{A picture of the four diamond sensors installed on the provisional beam pipe in the first commissioning phase (from February 2016 to June 2016).}
\label{beast}
\end{figure}

In the commissioning of SuperKEKB, two main phases are foreseen. In the first phase (BEAST II phase 1), from February 2016 to June 2016, each accelerator component, the beam injection systems and the two rings have been commissioned and vacuum scrubbing before the Belle II detector is rolled in has been performed.
Moreover a prototype detector system has been installed in order to provide measurements of the luminosity and background levels. These studies are needed in order to provide feedback to the accelerator during commissioning, to ensure a sufficiently low radiation level before the final Belle II detector is installed and to tune the simulation of the beam induced background. There are large uncertainties in the levels predicted by simulation, so that direct in situ measurements of the backgrounds are needed.
During phase 1 four diamonds (\ref{beast}) and the prototype readout electronics have been installed. 
The first results of the analysis of data from this first phase show a clear correlation with different beam conditions. The diamond sensors are very sensitive to the beam current value, the number of bunches and beam size.
In this first commissioning phase the radiation monitoring system (diamond sensors and the readout electronics) have been tested and the precision observed, 0.5 nA on the shortest 10$\mu s$ time scale, is reliable for fast and slow aborts for the next two commissioning phases.

\section{Conclusions}
The goal of the SuperKEKB collider is to provide 40 times higher instantaneous luminosity and 50 times higher integrated luminosity than that of the KEKB collider. As a consequence of the luminosity increase, severe beam- induced backgrounds and radiation doses are expected. A crucial aspect is the monitoring of both the radiation dose accumulated throughout the life of the experiment and the instantaneous radiation rate. A radiation monitoring and beam abort system based on twenty single-crystal diamond sensors will be adopted. The response of diamond sensors as dosimeters can be influenced by several issues related to the presence of traps inside the bulk and the properties of the diamond-electrode interface.
A complete characterisation of each sensor is then needed. 
Four diamond sensors (three single-crystalline and one poly-crystalline) have been calibrated at INFN-Trieste laboratories and then installed for the first SuperKEKB commissioning phase. In this phase the radiation monitoring system (diamond sensors and the readout electronics) have been tested and the precision observed, 0.5 nA on the shortest 10$\mu s$ time scale, is reliable for fast and slow aborts for the next two commissioning phases.

% that's all folks
\end{document}